\documentclass[11pt]{llncs}
\usepackage[T1]{fontenc}
\usepackage{graphicx}
\usepackage{amsmath}
\usepackage{amsfonts}
\usepackage{setspace}
\usepackage{amssymb}
\ifx\pdfoutput\@undefined\usepackage[usenames,dvips]{color}
\else\usepackage[usenames,dvipsnames]{color}
\IfFileExists{pdfcolmk.sty}{\usepackage{pdfcolmk}}{} 
\fi
\usepackage[plainpages=false,pdfpagelabels,pagebackref=false,naturalnames=true,hyperindex=true,pdftitle={From Computer Runtimes to the Length of Proofs},pdfauthor={Hector Zenil}]{hyperref}
\hypersetup{colorlinks=true,
urlcolor=Cerulean,linkcolor=BrickRed,citecolor=RoyalBlue,a4paper,
  pdfpagemode=None,
  pdfstartview=FitH}
\usepackage[all]{hypcap}

\begin{document}

\title{Computer Runtimes and the Length of Proofs\\ \normalsize{On an Algorithmic Probabilistic Application to\\Waiting Times in Automatic Theorem Proving}}
\titlerunning{From Computer Runtimes to Proof Lengths}
\toctitle{An Algorithmic Probabilistic Application to Theorem Proving Waiting Times}
\author{Hector Zenil\\ \institute{Department of Computer Science, University of Sheffield, UK\\ and Special Projects, Wolfram Research, Inc. USA.\\
\email{h.zenil@sheffield.ac.uk}}}
\date{}

\maketitle

\begin{abstract}

This paper is an experimental exploration of the relationship between the runtimes of Turing machines and the length of proofs in formal axiomatic systems. We compare the number of halting Turing machines of a given size to the number of provable theorems of first-order logic of a given size, and the runtime of the longest-running Turing machine of a given size to the proof length of the most-difficult-to-prove theorem of a given size. It is suggested that theorem provers are subject to the same non-linear tradeoff between time and size as computer programs are, affording the possibility of determining optimal timeouts and waiting times in automatic theorem proving. I provide the
statistics for some small choices of parameters for both of these systems.\\


\end{abstract}

\section{Introduction}

While profound connections between computer programs and mathematical proofs have been studied and are known (e.g. the Curry-Howard correspondence), little has been done to connect the two fields at the level of empirical practice. We present an experimental approach to the question of optimal proving times for automatic theorem provers, which bears out Calude and Stay's theoretical findings that programs either stop quickly or never halt \cite{calude}. 

Working with self-delimiting programs, that is, programs that are not the beginning of any other valid programs, Chaitin defined the complexity of the runtime of a program which eventually halts that we cannot effectively compute \cite{chaitin}, and Calude and Stay have recently proven \cite{calude} that even though short programs can run for a very long time, long programs are the scarcest because most of them will stop rather quickly---if they ever do---depending on their length. Thus, the probability of a machine halting decreases the longer it takes to halt, if it ever does. 

Just as Calude and Stay suggest that most Turing machines are fully determined qua termination by a small number of computational steps, and that the error margin drops drastically, in \cite{joosten} we have also shown that Turing machines  are fully determined qua extensionality by a small number of initial input values (a theoretical value for the error margin has yet to be determined but the very few data points that we could generate suggest to follow at least a polynomial distribution).

We undertake an experimental approach to the runtimes of deterministic Turing machines up to three states and two symbols in connection, and empirical evidence, to Calude and Stay's theoretical results. Then we undertake the same experimental approach to formulas of predicate calculus, in order to find some (if any) evidence in favour of a possible similar non-linear phenomenon in the distribution of proof lengths of (dis)proven theorems in random axiom systems  and Turing machines.

Traditional intuition might make one think this an ill-fated approach. On the one hand because undecidability would interfere in any such experimental attempt, and on the other hand, because small systems may say more about design choices than about important results. Even though possible limiting effects may appear right away one can limitedly circumvent these limits (as the Busy Beaver problem does) in an effort tantamount to other interesting experiments including some of Calude's own interest \cite{caludeomega} or of my own \cite{delahayezenil}, this latter providing useful  applications for the evaluation of the algorithmic complexity of short strings difficult to calculate with the other alternative (lossless compression algorithms). With the intuition one gets from studying small systems (see \cite{wolfram}), it seems worth it and insightful to undertake these kind of experiments.

\subsection{The halting problem}

The Halting Problem for Turing machines involves deciding whether an arbitrary Turing machine $M$ eventually halts on an arbitrary input $x$. One can ask whether there is a Turing machine halt $M$ which, given code $(M)$ and the input $x$, eventually stops and produces 1 if $M(x)$ stops, and 0 if $M(x)$ does not stop. Turing's seminal result states that this problem cannot be solved by any Turing machine, i.e. there is no such halt $M$. Halting can be recognized by simply running the machine in question; the main difficulty is to detect non-halting machines.

 Since many real-world problems arising in the fields of compiler optimization, automatized software engineering, formal proof systems, and so forth are deeply connected to the halting problem, there is an interest in understanding the problem in order to translate theoretical results into practical applications. 

In \cite{calude}, it was observed that for any computable probability distribution, most long times are effectively rare, so that at the limit they all had the same behavior regardless of the choice of distribution. They proved that the exact time at which a program stops is not too complicated algorithmically. It is (algorithmically) non-random because most programs either stop `quickly' or never halt. Since non-random times are (effectively) rare, according to Calude and Stay, the density of times at which an $N$-bit program can stop decreases quickly.

\section{The Busy Beaver problem}

There are $(4n+2)^{2n}$  possible $(n,2)$ deterministic Turing machines with $n$ states and 2 symbols. We denote by $(n,m)$ the class (or space) of all $n$-state $m$-symbol Turing machines having a bidirectional tape and remaining on the same cell when entering the (additional to $n$) halting state. Among the machines that halt, there are some that print more 1s on their output tapes than any other Turing machines of the same size, and some that reach a maximum number of steps upon halting.

 If $\sigma_T$ is the number of 1s on the tape of a Turing machine $T$ upon halting, then: $\sum(n)=\max{\{\sigma_T : T\in(n,2) \normalsize{\textbf{ }T(n)\textbf{ }halts}\}}$ with $n$ the number of states of the Turing machine.

If $t_T$ is the number of steps that a machine $T$ takes upon halting, then $S(n)=\max{\{t_T : T\in(n,2) \normalsize{\textbf{ }T(n)\textbf{ }halts}\}}$ with $n$ the number of states of the Turing machine.

$\sum(n)$ and $S(n)$ are noncomputable functions \cite{rado} by reduction to the halting problem. Yet values are known for $(n,2)$ with $n \leq 4$. The solution for $(n,2)$ with $n<3$ is trivial; the process leading to the solution in $(3,2)$ is discussed by Lin and Rado \cite{lin}; and the process leading to the solution in $(4,2)$ is discussed in \cite{brady}.

\subsubsection{Solving the halting problem for small machines}

 It is easy to see that $\sum(1)=1$ and $\sum(2)=4$. Lin and Rado \cite{rado} proved $\sum(3)=6$ and  Brady \cite{brady} that $\sum(4)=13$. The exact known values for $S$ are $S(1)=1$, $S(2)=6$, $S(3)=21$, $S(4)=107$. These Busy Beaver values are for 2-symbol Turing machines.
 
These numerical values of the Busy Beaver functions have been calculated by a combination of techniques, notably the exhaustive simulation of a reduced number of non-equivalent Turing machines, as it turns out that many can be decided (e.g. evident loops, etc) and because the number of cases is small enough one can either analyse case by case or actually run the machines and analyse their behaviour until deciding whether it halts or not. This is evidently possible because of the relatively small number of Turing machines with up to the number of states for for which the values of the Busy Beaver functions are known.

A program showing the evolution of all known Busy Beaver machines developed by this paper's authors is available online \cite{busyhz}. The formalism followed in this paper is the same as the one originally described and followed for the Busy Beaver problem as introduced by Rado \cite{rado}.

It is worth noting that the Busy Beaver problem is defined for Turing machines with initial empty tapes, and Turing machines studied in this paper are all provided with an initially empty tapes too. Turing universality tells us, however, that for every Turing machine with an arbitrary input there is a Turing machine with empty input computing the same function, hence Turing machines with empty tapes cover all possible cases (the translation may only result in some extra states).

\section{Halting and runtime distributions}

Calude and Stay showed that ``long-running" Turing machines can only halt at non-random times; the density of non-random times near $n$ is about $1/n$.  ``Long-running" means that if we have a universal Turing machine $U$ and machine $M$ is implemented by a program $m$ for $U$ of length $n$, then $U(m)$ runs for more than $c \times 2^n$ steps, where $c$ is some uncomputable constant depending on $U$.

\subsection{Halting history of $(2,2)$ Turing machines}

We know that a machine halts if it enters the halting state before reaching the known Busy Beaver value $S(n)$. If it does not, then it never halts. The halting problem and the halting probability problem are closely related to the Busy Beaver problem in that a solution to any one of them would yield a solution to each of the others.

Consider the halting space of all $(2,2)$ Turing machines (with an extra halting state) provided with an empty tape. The table in Fig. \ref{rundist} shows the runtime distribution at which all machines in $(2,2)$ halt (or do not).

\begin{figure}[h!]
\begin{center}\label{rundist}
\noindent\(
\begin{array}{|c|c|c|}
\hline
 \text{$t$} & \text{$k_t$} & \text{$p(k_t)$} \\
 \hline
 - & 6544 & 0.65 \\
 1 & 2000 & 0.20 \\
 2 & 800 & 0.080 \\
 3 & 160 & 0.016 \\
 4 & 56 & 0.0056 \\
 5 & 362 & 0.036 \\
 6 & 78 & 0.0078\\
 \hline
\end{array}
\)
\caption{Runtime distribution at which all machines halt (those that don't are indicated by ``---''). Where $t$ is the number of steps, $k_t$ the number of machines that halted at $t$ (out of a total of $3456$ that halt), and $p(k_t)$ is the halting probability of a machine to halt (or not) in time $t$.}
\end{center}
\end{figure}

There are 10\,000 2-state, 2-symbol Turing machines (the 10\,000 figure comes simply from the formula giving the number of Turing machines with $n=2$ states $(4n+2)^{2n}$). No other Turing machine halts after $6$ steps (see Fig. \ref{rundist}) in $(2,2)$. Machines that never halt are $6544$ in number, representing around $.65$ of the total.

What we term a runtime space is the product of a class of $(n,m)$ Turing machines for fixed $n$ and $m$, where programs are uniformly distributed, and the time space, which is discrete, has a halting time mapped to a greyscale color (the lighter the color, the sooner it halted; white means the program never halted and red means it reached the Busy Beaver value $S(n)$). 

\begin{figure}[h!]
\begin{center}\label{haltingspectra}
\includegraphics[height=.33in]{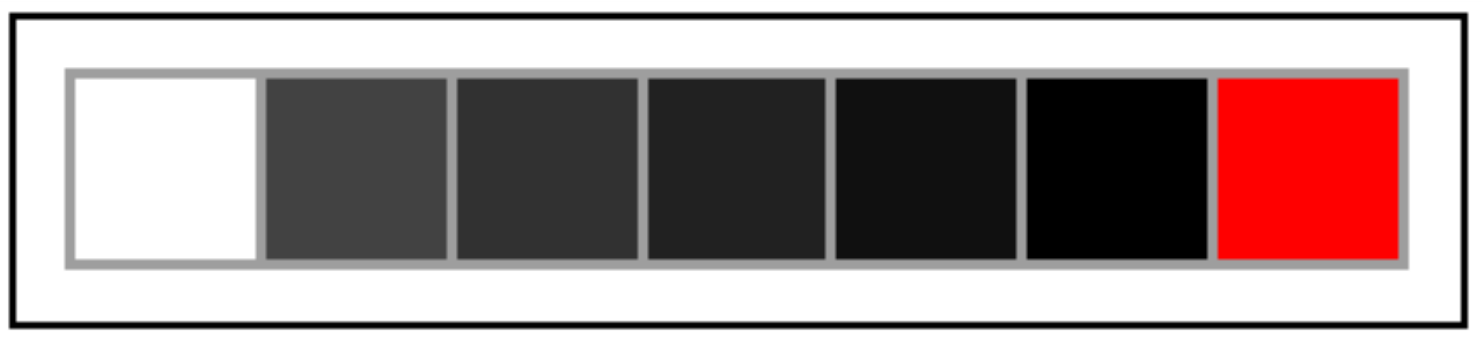}
\caption{Halting color mapping spectrum for Turing machines in $(2,2)$ (the last color is red, visible in the online and printed versions only).}
\end{center}
\end{figure}

\begin{figure}[h!]
\begin{center}\label{peanopacking1}
\includegraphics[height=3in]{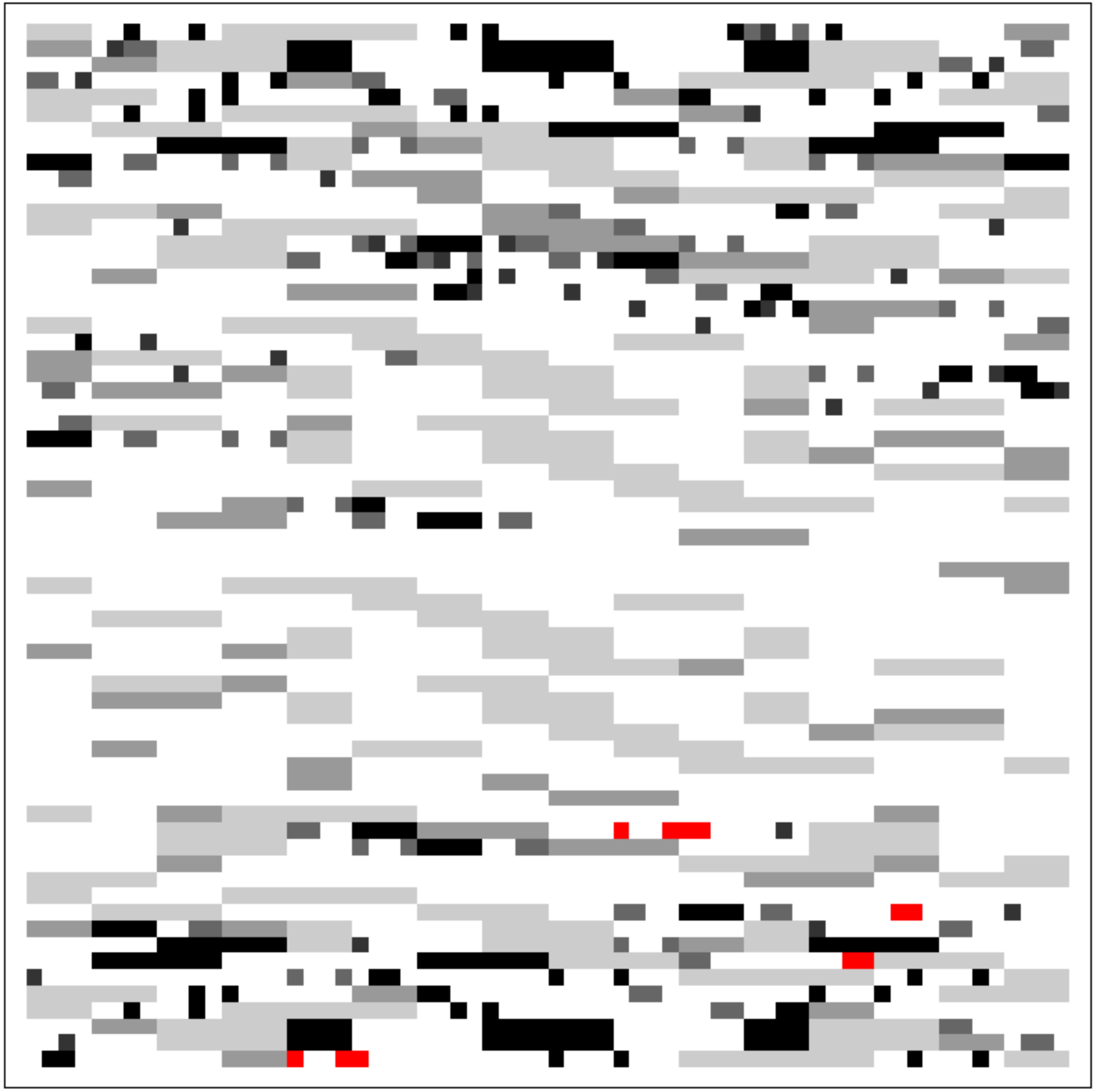}
\caption{Runtime distribution plot showing all the 10\,000 Turing machines in $(2,2)$ compressed in a Peano curve packing array (preserving the enumeration distance between machines). Some clusters may emerge due to the enumeration (e.g. terms involving transition rule parameters grouping Turing machines). The plot may look as if it had less than the necessary rows and columns to represent all the 10\,000 Turing machines, but that is a consequence of the Peano packing, each apparent pixel is in fact a small cluster of several machines.}
\end{center}
\end{figure}

Each point in Fig. \ref{peanopacking1} represents a Turing machine and as defined by the corresponding spectrum in Fig. \ref{haltingspectra}, the lighter the square the sooner it halted. White cells represent machines that don't halt. Red cells (only visible in the online and color printed versions) show the Busy Beaver machines (for this space, with runtime $S(2)=6$ steps). 

\begin{figure}[h!]
\begin{center}\label{accumulated}
\includegraphics[height=1.43in]{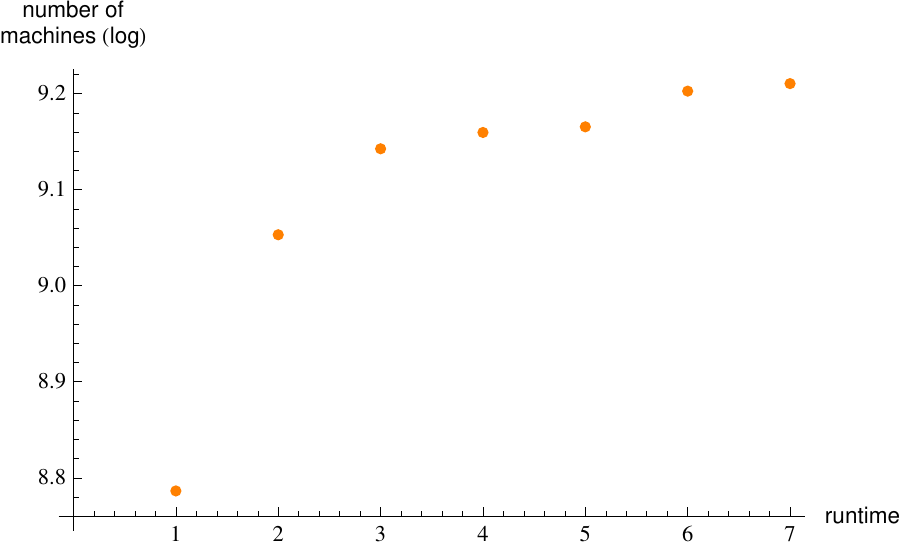}
\includegraphics[height=1.43in]{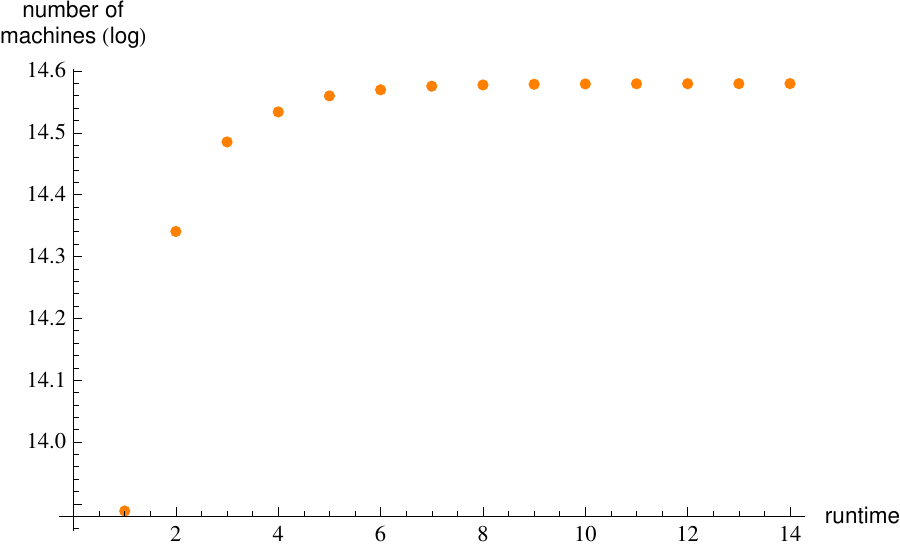}
\caption{Accumulated number of machines in $(2,2)$ (left) and $(3,2)$ (right) that halt step by step.}
\end{center}
\end{figure}

Among all the $3456$ Turing machines in $(2,2)$ that halt, .65 of them do so after the first step, $.2$ do so after the second, $.05$ after the third, and so on. In other words, $.57$ out of the $3456$ $(2,2)$ Turing machines that halted did so at the first step, $.81$ halted before or by the second step at the latest, $.84$ before or by the third step at the latest, and so on (see Fig. \ref{accumulated}).

\subsection{Halting history of $(3,2)$ Turing machines}

\begin{figure}[h!]
\begin{center}\label{TM32}
\noindent\(
\begin{array}{|c|c|c|c|}
\hline
 \text{$t$} & \text{$k_t$} & \text{$100\times2^{14-t}$} & \text{$p(k_t)$} \\
\hline
 - & 5382624 & & 0.71 \\
 1 & 1075648 & 819200 & 0.14 \\
 2 & 614656 & 409600 & 0.082 \\
 3 & 263424 & 204800 & 0.035 \\
 4 & 97216 & 102400 & 0.013 \\
 5 & 53760 & 51200 & 0.0071 \\
 6 & 20800 & 25600 & 0.0028 \\
 7 & 12512 & 12800 & 0.0017 \\
 8 & 4264 & 6400 & 0.00057 \\
 9 & 2424 & 3200 & 0.00032 \\
 10 & 1064 & 1600 & 0.00014 \\
 11 & 536 & 800 & 0.000071 \\
 12 & 304 & 400 & 0.000040 \\
 13 & 176 & 200 & 0.000023 \\
 14 & 128 & 100 & 0.000017 \\
 \hline
\end{array}
\)
\caption{Where $t$ is the number of steps, $k_t$ the number of machines that halted at $t$, and $p(k_t)$ is the halting probability calculated from $t$ and $k_t$. $100 \times 2^{14-t}$ is a good fit to the limit behavior as a function relating runtimes and the number of Turing machines halting at a certain runtime for the 14 runtimes at which Turing machines halt.}
\end{center}
\end{figure}

\begin{figure}[h!]
\begin{center}
\includegraphics[height=2in]{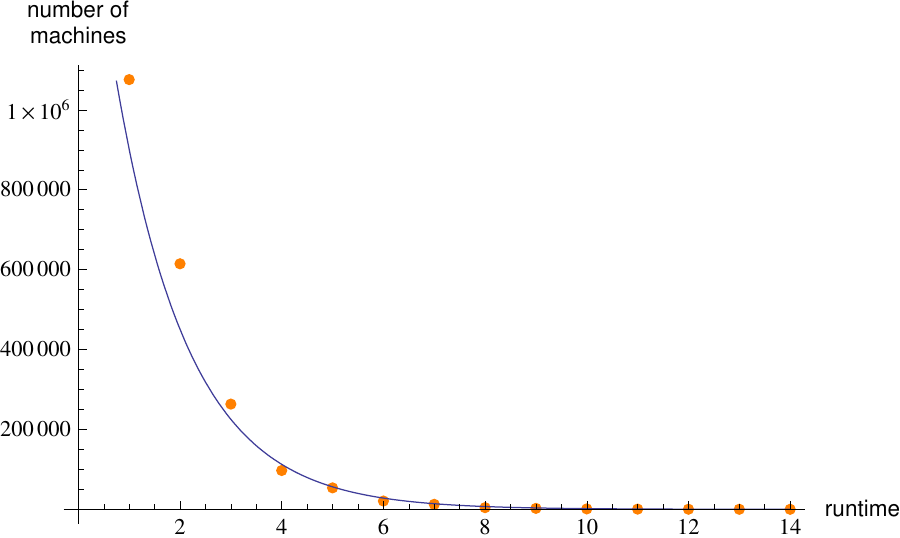}
\caption{Number of machines in $(3,2)$ that halt step by step versus $100 \times 2^{14-t}$ (dark line (blue in color version)).}
\end{center}
\end{figure}

Interesting output distribution facts:

\begin{itemize}
\item Out of 7\,529\,536 machines only 2\,146\,912 halt.
\item There are 5\,382\,624 machines that do not halt.
\item Those machines that halt only produce $126$ different output strings, with the largest being $6$ digits in length (the Busy Beavers).
\item Exactly $.2$ of the Turing machines produce a $0$ or a $1$ as output.
\end{itemize}

The fact that the figures are mostly white and lightly colored is an indicator of the sparsity of non-halting or quickly-halting machines.

\begin{figure}[h!]
\begin{center}
\includegraphics[height=.4in]{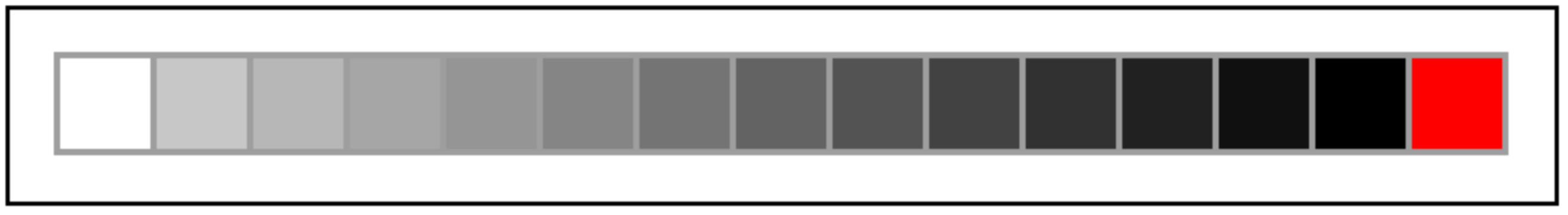}
\caption{Halting spectrum for $(3,2)$. Last color in the spectrum is red (only visible in the online and color printed versions)}
\end{center}
\end{figure}

\begin{figure}[h!]
\label{deepfield}
\begin{center}
\includegraphics[height=3in]{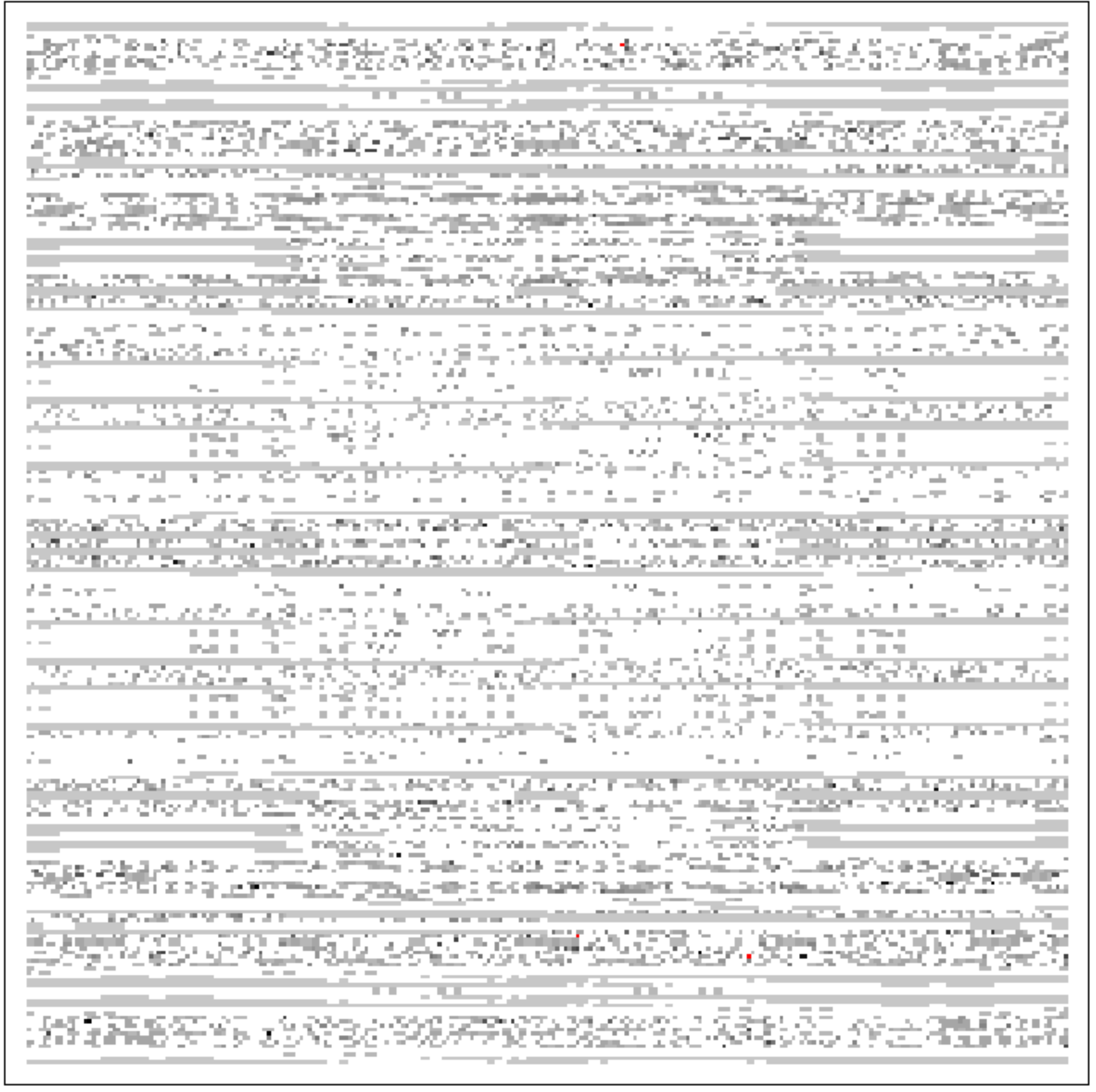}
\caption{Runtime deep field of a segment of runtimes from the 7\,529\,536 Turing machines in $(3,2)$. The $(3,2)$ Busy Beavers are barely visible as isolated red points (online and color printed versions only).}
\end{center}
\end{figure}

\begin{figure}[h!]
\begin{center}
\includegraphics[height=2.7in]{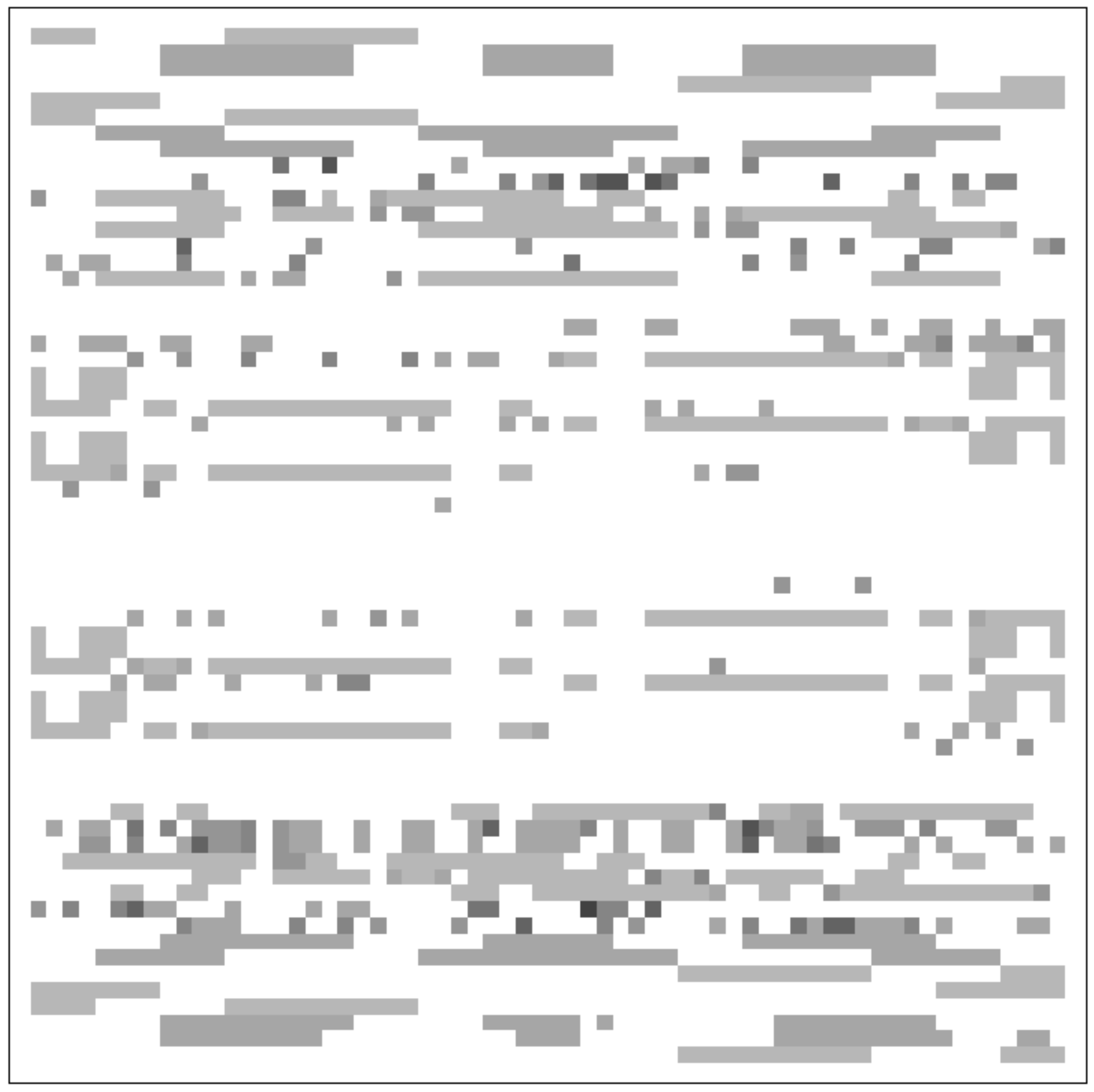}
\caption{This is what a typical random part of the runtime deep field looks like after a $10\times$ zoom from a 10th. square area of the original (Fig. \ref{deepfield}) image.}
\end{center}
\end{figure}

\newpage

\section{G\"odel meets Turing in the computational universe}

Inspired by \cite{wolfram} where Wolfram undertakes an exhaustive investigation of the space of propositional logic formulas, I extended his ideas to investigate the space of first order logic. The extension wasn't trivial, among other reasons because unlike propositional calculus, predicate calculus is undecidable, meaning that one may come across cases where formulas (or their negations) are not proven or disproven in an axiom system of first order logic.

 Proof lengths are, of course, not bounded, or one would be able to decide whether a formula in an axiom system can be proven or not if it has reached a limit. Frequency of proof lengths for randomly generated formulas, however, can be studied and analyzed. Frequency distributions of (dis)proven formulas turn out to follow a similar distribution to those of randomly generated computer programs, in which most programs, just as we found for formulas, halt (or are (dis)proven) quickly, with their number diminishing fast over time. When I met Cris Calude and became acquainted with his fascinating work, including a recent collaboration with Michael Stay on the distribution of halting times of random computer programs \cite{calude}, it prompted me to seek connections with these other findings---persuaded as I was of the strong connections known to exist between computation and proof theory---and to undertake an empirical investigation of both the halting runtimes of Turing machines that Calude and Stay had calculated theoretically, and the lengths of proofs found by automatic theorem provers.
 
It follows from Chaitin \cite{chaitin} and Calude and Stay \cite{calude} that to (dis)prove a formula in an axiom system one only needs to check up to the runtime for which the Turing machine encoding the proof no longer halts. Busy Beavers, as used in the previous section, are therefore relevant to automatic theorem proving because they provide an upper bound on the length of proofs. One only needs to run the computer to (dis)prove the formula up to the Busy Beaver value of the size of the Turing machine, and if it cannot be proven by then then it is undecidable for that axiom system. Moreover, Calude and Stay's work may then suggest that chances of proving a formula should decrease over time, or that if a formula can be (dis)proven it will likely do so early in time rather than later meaning that one can set an optimal time for a given provability certainty goal.

\subsection{Computer runtimes and lengths of proofs}

Optimal proving times are relevant because, on the one hand, they may allow one to set a maximum waiting time, given that proofs may never arrive if a theorem is undecidable in an axiom system, but also because one would know how long to wait before giving up with a certain degree of certainty of provability. If one had a goal (say to prove a fraction of .90 of a set of formulas) one could calculate an optimal timeout and a maximum waiting time, taking advantage of the fact that in the case of theorem provers running on digital computers, there is a correspondence between runtime and proof length. The numbers involved are so large and grow so fast because of the combinatoric explosion (in the number of formulas as well as the number of Turing machines). We were only able to explore the tip of the iceberg of the space of all possible first-order formulas, but with interesting and encouraging results nonetheless.

\subsection{Enumerating and generating predicate calculus axiom systems with equality}
\label{intro}

A number of sound and complete calculi have been developed enabling fully automated theorem provers for first-order logic. Equational logic is quite simple, and yet powerful \cite{baumgartner}. Its atomic formulas are equations, making it very easy to encode and deal with. In our formalism, terms are first-order formulas built from variables and constants using function symbols. Equalities of the form $lhs = rhs$ are the atomic formulas in our language, where $lhs$ and $rhs$ are terms. One can represent most mathematical axiom systems and theorems in equational form, so it is expressively very rich. A logical system which possesses an explicitly stated set of axioms from which theorems can be derived is an axiomatic system.

 In predicate calculus, a formula is in \emph{prenex} normal form if it can be written as a string of quantifiers followed by a quantifier-free part. All first-order well-formed formulas (hereafter simply `formulas') are logically equivalent to some formula in \emph{prenex} normal form. \emph{Skolemization} is a way of removing existential quantifiers from a formula. Variables bound by existential quantifiers which are not within the scope of universal quantifiers can simply be replaced by the appropriate constants. Both will be used in order to enumerate all possible quantified axioms and formulas of first order logic.

 All equational formulas can be represented with two binary operators $f$ and $p$, where $p$ is a pairing function and $f$ is an indexing operator (any possible binary function). The first parameter of $f$ will be a constant determining its index, while the second is any other term (variable, constant, $f$ itself or $p$).

 When the existential quantifier is inside a universal quantifier, the bound variable must be replaced by a Skolem function of the variables bound by universal quantifiers. We can then specify any constant using a formula of the form: $\forall_a \forall_b f(a,a)=f(b,b)$. And the $i-$th constant can be defined in terms of $f$ and $p$ recursively as follows:

\begin{center}
\noindent\text{c(0)=p(f(a, a), f(a, a))}\\
\noindent\text{c(n+1)=p(f(a, a), c(n))}
\end{center}

Or in a single $Mathematica$ expression:

\begin{center}
\noindent\text{Nest[p[f[a,a],\#]\&,p[a,a],i]}
\end{center}

To represent all possible functions one can combine both $f$ and $p$. For instance, $f(c(i),p(c(i),x))$ is the expression representing the $i$-th function (the function with index $i$) of $x$. This assumes that there are an infinite number of individuals in the most general case. Notice that $x$ may be a list built from pairs. 

Formulas were enumerated and generated by the number of variables and constants on both sides of the equality. There are no formulas of length 1, simply because an equality requires at least 2 terms on each side. Finally, all single axioms were arranged by length. The length of an equational formula is the sum of the bound variables on both sides of the equality. Axiom systems are simply all the possible subsets over the formulas of fixed length. Applying this operation makes the number of axiom systems to grow exponentially, so we were able to proceed exhaustively only up to 3 bound variables formulas and to generate a sample of 1000 axiom systems only (an initial segment) for 4 bound variables formulas. An automatic theorem prover was fed with all 4 bound variable single formulas as its proving goal for each of the generated axiom system, producing almost $10\times10^3$ proofs. Among the initial 1000 axiom systems, 607 were used only, as they were proven to be consistent (no axiom was the negation of any other) and independent (no axiom could be derived from the others).

An example of a formula with 3 bound variables is: $\forall_{x_1} \forall_{x_2} \forall_{x_3}, x_1 = f(f(x_2, x_3), x_1)$ and with four: $\forall_{x_1} \forall_{x_2} \forall_{x_3} \forall_{x_4}, x_1 = p(f(x_2, x_3), x_4)$. An example of an axiom system consisting of 2 axioms each with 2 bound variables is: $\forall_{x_1} \forall_{x_2}, x_1 = f(x_2, x_1) \wedge \forall_{x_1} \forall_{x_2}, x_1 = p(x_1, x_2)$. Notice that one does not need to further compose $f$ with $p$ or $p$ with $f$ in order to produce other possible formulas, because $f$ is a general function with an index as first parameter and any term as second parameter which can be $p$ or $f$ itself, without the need of infinitely nesting each into the other in order to reach other possible constructions.

\subsection{Experimental setting}

The project was undertaken using \emph{Mathematica}'s built-in implementation of the well known and award-winning theorem prover \textit{Waldmeister}\footnote{\url{http://www.mpi-inf.mpg.de/~hillen/waldmeister/} (August, 2011).}. \emph{Waldmeister} returns True after evaluating an expression in \emph{Mathematica} if it can prove the conclusions from the given axioms, and False if it can prove that the conclusions do not follow from the axioms. If it cannot prove either, it returns Unevaluated.

 The axiom systems generated---as described in section \ref{intro}---were first checked for logical consistency and internal axiom independence, these being two of the most important qualities of conventional mathematical axiom systems. $A$ is said to be consistent if no theorem and its negation can be derived from $A$. On the other hand, if $A$ is an axiom system and $a\in A$, then a is considered independent in $A$, or an independent axiom of $A$ if $a$ cannot be derived from $A-\{a\}$. As with any axiomatic system, we want this axiomatic system to be minimal, i.e. to contain no superfluous axiom. From this point on, only consistent axiom systems were taken into account. \\

Miscellaneous interesting first results:

\begin{itemize}
\item It was found that only $.01$ out of a total of $490$ axiomatic systems with 1 or 2 axioms of length up to 3 bound variables were non-independent, i.e. one of its members could be derived from a combination of the others. 
\item All the $29$ axiomatic systems of length 3 with 2 or more axioms were independent. This could be explained by the way in which the axiomatic systems were enumerated, because axioms closer to each other in the enumeration seem to have a better chance of being derived from each other. The condition of being a theorem or an axiom is evidently an arbitrary convention.
\item The number of consistent axiom systems of length 3 was only $.0342$ percent of a total of $1024$ initial axiomatic systems. 
\item In the case of axiom systems of length 4 (composed by formulas of that size), $.607$ of them were found to be consistent. This may be interpreted in two different ways: that even when the complexity of the axiom systems grows, the overall inconsistency does not increase, or else that the process only unveils the tip of the iceberg, where they are consistent chiefly due to their simplicity (both in terms of number of axioms per axiom system and the length of the axioms themselves, thereby reducing the possible number of clashes).
\end{itemize}

\subsection{Distribution of proof lengths}

The relation between the length of the formulas and the optimal runtime limit is of particular utility when no upper bound is known (or possible), when, for example, there are non-provable formulas for which longer runtimes will not make any difference---which, as verified herein, would cover a negligible number of cases. 

A total of 89\,145 formulas out of the 97\,727 with at most 4 variables were proven to be theorems (or their negations) after a single step. One can call such a theorem \emph{trivial} simply because its proof, requiring only 1 step, can be accomplished with an axiom, therefore itself being an axiom. The proof length ($t$) distribution (in percentage) of formulas with up to 4 variables is as shown in Fig. \ref{theo}.

\begin{figure}[h!]
\begin{center}\label{theo}
\noindent\(
\begin{array}{|c|c|c|}
\hline
 \text{$t$} & \text{$k_t$} & \text{$p(k_t)$} \\
\hline
 1 & 89145 & 91.2184 \\
 2 & 2311 & 2.36475 \\
 3 & 473 & 0.484001 \\
 4 & 931 & 0.952654 \\
 5 & 928 & 0.949584 \\
 6 & 426 & 0.435908 \\
 7 & 577 & 0.59042 \\
 8 & 834 & 0.853398 \\
 9 & 1344 & 1.37526 \\
 10 & 294 & 0.300838 \\
 11 & 186 & 0.190326 \\
 12 & 206 & 0.210791 \\
 13 & 44 & 0.0450234 \\
 14 & 15 & 0.0153489 \\
 15 & 7 & 0.00716281 \\
 16 & 2 & 0.00204652 \\
 17 & 4 & 0.00409303\\
 \hline
\end{array}
\)
\caption{Proof length ($t$) distribution (in percentage) of formulas with up to 4 variables.}
\end{center}
\end{figure}

Proof length distribution of (dis)proven theorems. Where $t$ is the number of steps the theorem prover has taken to produce the proof, $k_t$ the number of machines that halted at $t$, and $p(k_t)$ is the halting probability of having (dis)proven $k$ theorems in time $t$ from which one can build a probability distribution $p(k_t)$.

%\begin{figure}[h!]
%\begin{center}
%\includegraphics[height=2in]{RuntimesOfProofsAndPrograms-Zenil_gr10.eps}
%\caption{Finer inspection by looking at non-trivial proofs (i.e. proofs that take at least 2 steps or more).}
%\end{center}
%\end{figure}

It is worth noting that the behavior of \ref{theo} graph resembles the first case of $(2,2)$ Turing machines, where the number of machines that halted was not strictly decreasing (unlike $(3,2)$ that was monotonically decreasing). 

\begin{figure}[h!]
\begin{center}
\includegraphics[height=1.9in]{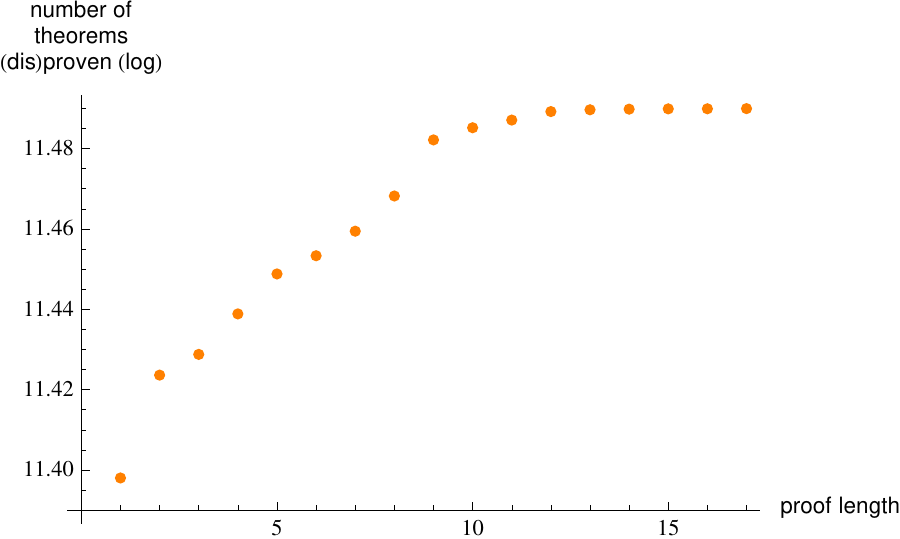}
\caption{Accumulated number of theorems (dis)proven step by step.}
\end{center}
\end{figure}

Already $0.912$ out of the total number of theorems are proven by the very first step, with that number dropping as the total is approached. From the distribution it follows that going beyond the 7th. step to the 17 steps that require the longest proofs only adds $.012$ new (dis)proven formulas to the total. Summary of proving times:

\begin{itemize}
\item A total of $89\,145$ formulas out of $97\,727$ were immediately proved (or disproved) after the first step (i.e. $91.21\%$). 
\item $95.96$ were proven after $5$ steps, and $96\,969$ formulas were proven after $9$ steps (which is almost half of the $17$ maximum number of steps reached by the formulas with $4$ bound variables). That is, $99.22\%$ of the total. 
\item Letting the theorem prover run up to $17$ steps only generates $758$ new proofs, that is only $0.77\%$ of the total.
\end{itemize}

%\begin{figure}[h!]
%\begin{center}
%\includegraphics[height=.5in]{proofspectra}
%\caption{Proof length color mapping spectra.}
%\end{center}
%\end{figure}

\begin{figure}[h!]
\label{truthspace}
\begin{center}
\includegraphics[height=1.42in]{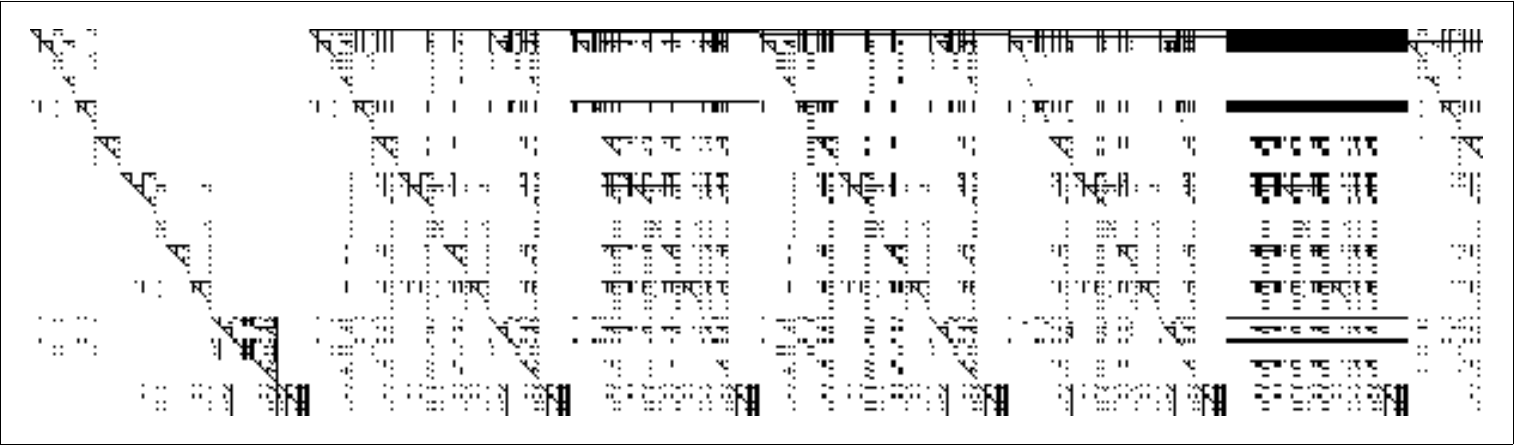}
\caption{Truth space of 97\,727 proofs from the 607 consistent and independent axiom systems ($x$ axis) against 161 formulas ($y$ axis) from formulas with 4 bound variables. Every dot is a proof, a black square indicates that a particular theorem holds in a particular axiom system (which explains the diagonal, among other patterns) and white means the formula was proven to be false in the corresponding axiom system (i.e. the negation is a theorem). No undecidable candidate was found.}
\end{center}
\end{figure}

\begin{figure}[h!]
\begin{center}
\includegraphics[height=.5in]{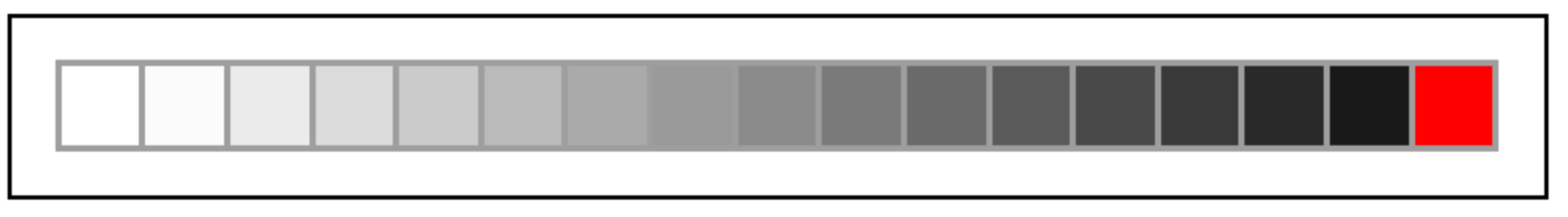}
\caption{Color mapping spectrum for proofs of length 4.}
\end{center}
\end{figure}

\begin{figure}[h!]
\label{godelspace}
\begin{center}
\includegraphics[height=1.47in]{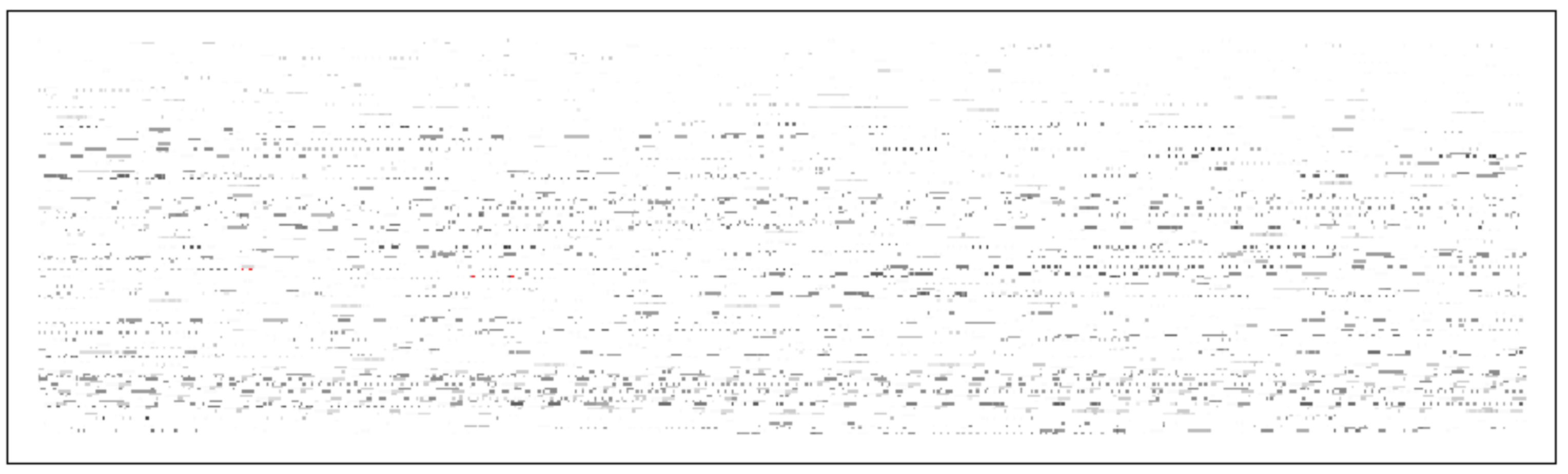}
\caption{Proof length deep field plot from the 97\,727 formulas of up to 4 variables. \emph{formula Busy Beavers} are barely visible as isolated red points (online and color printed versions only). Points are arranged as in \ref{truthspace}.}
\end{center}
\end{figure}

As for Turing machines (see Fig. \ref{deepfield}), the space of proof lengths (Fig. \ref{godelspace}) is mostly white and lightly colored as an indicator of the sparsity of long proof lengths given that most formulas are (dis)proven very quickly, suggesting that the distribution of proof lengths follows the distribution of program runtimes.

\begin{figure}[h]
\label{limit2}
\begin{center}
\noindent\(
\begin{array}{|c|c|c|}
\hline
 \text{runtime} & \text{(dis)proven fraction} & \text{$f(t)=1/2^t$} \\
 \text{$t$} & \text{of theorems $p(t)$} & \text{(first significant digit)}\\
\hline
 1 & 0.9 & 0.5 \\
 2 & 0.02 & 0.2 \\
 3 & 0.005 & 0.1 \\
 4 & 0.01 & 0.06 \\
 5 & 0.009 & 0.03 \\
 6 & 0.004 & 0.02 \\
 7 & 0.006 & 0.008 \\
 8 & 0.009 & 0.004 \\
 9 & 0.01 & 0.002 \\
 10 & 0.003 & 0.001 \\
 11 & 0.002 & 0.0005 \\
 12 & 0.002 & 0.0002 \\
 13 & 0.0005 & 0.0001 \\
 14 & 0.0002 & 0.00006 \\
 15 & 0.00007 & 0.00003 \\
 16 & 0.00002 & 0.00002\\
 17 & 0.00001 & 0.00001\\
\hline
\end{array}
\)
\end{center}
\caption{Runtime distribution at which all machines halt (those that don't are indicated by ``---''). Where $t$ is the number of steps, $k_t$ the number of machines that halted at $t$ (out of a total of $3456$ that halt), and $p(k_t)$ is the halting probability calculated from $t$ and $k_t$.}
\end{figure}

%\begin{figure}[h]
%\begin{center}
%\includegraphics[height=1.75in]{RuntimesOfProofsAndPrograms-Zenil_gr15.eps}
%\caption{The scatter log plot shows the proof lengths (dots) versus the absolute values of $1/2^n$ (line).}
%\end{center}
%\end{figure}

\section{Timeouts and optimal waiting times}

As for Busy Beaver Turing machines, the values of which depend on the size of the Turing machines (states and symbols), proof lengths depend on the length of the formulas. One can define \emph{Busy Beaver formulas} (the values of which will be denoted by $fBB(n)$) as the formulas for which an automatic theorem prover takes more time to (dis)prove whether a theorem is decidable, or to produce the longest proof, among all the formulas of a fixed length. Unlike Turing machines, however, the size of a formula can take many forms, and may depend on the number of bound variables (as was the case in the experiments undertaken here), the number of logical operators or the number of symbols in general. It also depends on the formalism, just as Busy Beavers depend on the formalism used by Rado \cite{rado}. Following the analogy, the values of $fBB(n)$ would therefore work in a similar way and may be used just as Busy Beaver Turing machine values are currently used---for defining maximum runtimes and maximum output lengths for (small) Turing machines, saving time once an upper limit is known. The exact relation would also save considerable computational resources in automatic theorem proving. 

As explained before, the theoretical algorithmic analysis in \cite{calude} indicates that a program that has not stopped after running for a long time has smaller and smaller chances of eventually stopping, so the longer the time $t$ the more unlikely the program is to halt.  Calude and Stay's results can be interpreted as follows: most Turing machines are fully determined qua termination by a small number of computational steps, and the error margin upon betting that a Turing machine will halt drops exponentially. Because proofs are programs for automatic theorem prover and one can connect this interpretation to the probability of a formula to be (dis)proven in an axiom system with a confidence error margin to be proven dropping fast.

Let the ``optimal timeout'' be the number of steps for which a fraction of formulas from a set of fixed length is (dis)proven. Evidently, proving time is asymptotically optimal, in the sense that the closest to the maximum runtime (the Busy Beaver formula values), the greatest the fraction of (dis)proven formulas. An optimal time $OPTime$ for a given goal $\gamma$ implies that upon $t$ one has reached a fraction $\gamma$ of (dis)proved formulas. Thus $OPTime(n,\gamma) = \min\{t(n) : | \alpha_{t(n)} | = \gamma\}$, where $n$ is the length of the set of formulas, $\gamma$ the desired fraction of (dis)proved formulas and $| \alpha |$ the number of formulas proven at time $t(n) \geq 0$. Obviously $0< OPTime(n) \leq fBB(n)$ for each time $t>0$, and $OPTime(n)= fBB(n)$ if $\gamma=1$, that is, if the fraction of formulas to be (dis)proved is 1 (i.e. if the goal is to (dis)prove all the formulas of a fixed length).

 Just as with Busy Beavers, the exact value of $OPTime(n)$ is uncomputable and unpredictable in general, but one can approach it. For example, in our formalism, for 4 bound variables it can be calculated from the probability distribution in \ref{limit2}. One can ascertain, for example, that from a uniform distribution of randomly generated formulas, nearly .90 of the formulas will be proven after the first step. And that the number of new proofs from then on will rapidly drop as a function of the number of steps. The value of $OPTime(n)$ can also determine a timeout for single formulas, given a confidence expectation. Which is to say that a single formula has, for example, a .90 chance of being (dis)proven in the first step, and that it has diminishing possibilities, if any, of being (dis)proven thereafter. 
We think that the results are robust enough to model specifications of theorem provers, despite not being completely independent. We were able to verify the results using another very different theorem prover, the Automatic Proof Search or AProS \cite{sieg} for propositional logic and predicate calculus (the theorem prover deals, however, with all sorts of other classical and non-classical calculus). AProS uses the intercalation method to search for normal natural deduction proofs not requiring a language in which the atomic formulas are identities, unlike \emph{Waldmeister}. Notice that for this new case, the definition of the length of formulas was adjusted to the new framework, given that since the prover calculus does not require equality, no sense can be given to left or right hand sides. The set of randomly chosen operators used to generate formulas were the classic \emph{and}, \emph{or}, \emph{implies} and \emph{double implies}. AProS found proofs for .12 of the assertions (and for .353 of a set of assertions with no-double conditionals), out of a random choice of 1000 automatically generated predicate calculus assertions with up to 4 quantifiers, 3 general functions, 3 logical operators and 3 variables. The longest proof length (runtime) was 42 with an average proof length of 13, and a distribution very close to the one described by \emph{Waldmeister} using \emph{Mathematica}.

\section{Concluding remarks and further work}

A logically significant question concerns the structure of the theorems established. If significant structural features are uncovered, then one could generate randomly formulas of that structure and repeat the proof length and runtime distribution experiments. It would be quite interesting, if one could find, for example, systematic biases for different theorem provers and theorem proving techniques when deviating in distribution from each other.

One can continue the process of generalizing theoretical results from computer programs to proof lengths and seek the equivalent of Busy Beavers in sets of well defined proofs and theorem provers. Just as for larger Busy Beaver Turing machine values, the computer time and resources to explore much larger sets of proofs are out of reach. The experiments suggest that the statistics for theorem proving times from randomly generated formulas may follow a similar trend to the distribution of runtimes of random computer programs. And that when searching for proofs, appropriate timeouts can be set and optimal waiting times defined depending on the size of the formulas as it has been determined that runtimes depend on the size of machines. It is too soon, however, to declare any true resemblance and there are always dangers of extrapolating from the behavior of small systems.

\subsection*{Acknowledgments}

I am grateful to Cris Calude who encouraged me to publish these results in connection with his own work \cite{calude}. I am also indebted to Stephen Wolfram, Todd Rowland and Matthew Szudzik for their support and guidance during and after the 2005 NKS Summer School at Brown University, when I started this project as part of a 3-week Summer project and inspired by Stephen Wolfram's own work in \cite{wolfram}, intending to extend his results from propositional logic to predicate calculus. I am also grateful to J.-P. Delahaye with whom I've undertaken related research \cite{delahayezenil}, studying the output distribution of abstract computing machines. To Wilfried Sieg for his guidance and for introducing me to AProS, which I used to strengthen the experimental results in this paper while a visiting scholar at Carnegie Mellon, and to Jeremy Avigad who brought me to Carnegie Mellon. And to the anonymous referee. Any error or omission remains, of course, the sole responsibility of this author.

\bibliographystyle{splncs}

\end{document}